\documentclass[10pt,conference]{IEEEtran}\IEEEoverridecommandlockouts
\usepackage{epsfig,graphicx,subfigure,psfrag,amsmath,cases}
\usepackage{latexsym,amssymb,amsmath,epsfig,subfigure,algorithm,mathtools,empheq}
\usepackage{algorithmic}
\usepackage{color}
\usepackage{url}
\usepackage{scrtime}
\usepackage{array}

\author{\authorblockN{Derrick Wing Kwan Ng and Robert Schober\authorrefmark{1}\thanks{\authorrefmark{1}The author is also with the University of British Columbia. This work was supported in part by the AvH Professorship Program of the Alexander von Humboldt Foundation.}}
Friedrich-Alexander-University Erlangen-N\"urnberg (FAU), Germany\\
Email:  kwan@lnt.de,  schober@lnt.de\vspace*{-4mm}
}

\title{\vspace*{-2mm}Max-min Fair Wireless Energy Transfer for Secure Multiuser Communication Systems}

\date{\thistime,\,\today}

\newtheorem{Thm}{Theorem}
\newtheorem{Lem}{Lemma}

\newtheorem{proposition}{Proposition}

\DeclareMathOperator{\Tr}{\mathrm{Tr}}
\DeclareMathOperator{\Rank}{\mathrm{Rank}}

\DeclareMathOperator{\zero}{\mathbf{0}}
\DeclareMathOperator{\maxo}{maximize}

 \newcommand{\qed}{\hfill \ensuremath{\blacksquare}}
 
\newcommand{\abs}[1]{\lvert#1\rvert}
\newcommand{\norm}[1]{\lVert#1\rVert}

\newcolumntype{L}{>{\hspace*{-1.5mm}\arraybackslash}m{5cm}}
\newcolumntype{M}{>{\hspace*{-1.5mm}\arraybackslash}m{3cm}}

\begin{document}
\IEEEspecialpapernotice{(Invited Paper)}
\maketitle

\begin{abstract}
This paper considers max-min fairness for wireless energy transfer in a  downlink multiuser communication system. Our resource allocation design maximizes the minimum harvested energy among multiple  multiple-antenna energy harvesting receivers (potential eavesdroppers) while providing quality of service (QoS) for secure communication to multiple single-antenna information receivers. In particular,
the algorithm design is formulated as a non-convex
optimization problem which takes into account a minimum required signal-to-interference-plus-noise ratio (SINR) constraint at the information receivers and a constraint on the maximum tolerable channel capacity achieved by the energy harvesting receivers  for a given transmit power budget.
The proposed problem formulation exploits the dual use of  artificial
noise generation for facilitating efficient wireless energy transfer and secure communication. A  semidefinite programming (SDP) relaxation  approach is exploited to obtain  a global optimal solution of the considered  problem. Simulation results demonstrate the significant performance gain in harvested energy that is achieved by the proposed optimal scheme compared to two simple baseline schemes.
\end{abstract}

\renewcommand{\baselinestretch}{0.94}
\large\normalsize

\section{Introduction} \label{sect1}
Ubiquitous, secure, and high data rate communication  is a basic requirement for the next generation  wireless communication systems.
The rapid growth  of wireless
data traffic in the past decades has heightened the energy consumption in both transmitters and receivers. As a result, multiuser multiple-input multiple-output (MIMO) has been proposed in the literature for facilitating energy efficient wireless communication. Although the energy dissipation of the transmitters may be significantly reduced by multiuser MIMO technology,  mobile communication devices and sensor devices are still often
powered by batteries with limited energy storage capacity.  Frequently replacing the device batteries can be costly and inconvenient in difficult-to-access environments, or even infeasible for medical sensors embedded inside the human body. Hence, the limited lifetime of communication networks constitutes a major bottleneck in providing quality of service (QoS) to the end-users.

 Recently, energy harvesting based mobile communication system design has drawn significant interest from both academia and industry  since it enables self-sustainability of energy constrained wireless devices. Traditionally,  wind, solar, and biomass, etc. are the major sources for energy harvesting. Although these renewable energy sources are perpetual, their availability usually depends on location and climate which may not be suitable for mobile devices. On the other hand, wireless energy transfer technology,  which  allows receivers to scavenge energy from the ambient radio frequency (RF)  signals,   has attracted
much attention lately although the concept can be traced back to Nikola Tesla's work in  the early 20th century \cite{Pa:Nikola}.
There have been some preliminary applications of wireless energy transfer  such as wireless body area networks (WBAN) for biomedical implants, passive radio-frequency identification (RFID) systems, and wireless sensor networks. Indeed, the combination of RF energy harvesting and communication provides the possibility of simultaneous wireless information and power transfer (SWIPT) which imposes many new and interesting challenges for wireless communication engineers \cite{CN:Shannon_meets_tesla}--\nocite{JR:MIMO_WIPT,JR:WIP_receiver,
JR:Rui_zhang_power_splitting,
CN:WCNC_WIPT,JR:WIPT_fullpaper}\cite{JR:Kai_bin}. In \cite{CN:Shannon_meets_tesla}, the  trade-off between channel capacity and harvested energy was  studied for near-field communication over a frequency selective channel. In \cite{JR:MIMO_WIPT}, the authors investigated the performance limits of a three-node wireless MIMO broadcast channel for SWIPT. In particular, the tradeoffs between maximal information
rate versus energy transfer were characterized by the
boundary of a rate-energy (R-E) region. In \cite{JR:WIP_receiver},  a power splitting receiver and a separated receiver were proposed to realize concurrent information decoding and energy harvesting for narrow-band single-antenna communication systems. In \cite{JR:Rui_zhang_power_splitting}--\nocite{CN:WCNC_WIPT,JR:WIPT_fullpaper,JR:Kai_bin}\cite{JR:WIPT_fullpaper}, different resource allocation algorithms were proposed for improving the utilization of limited system resources. Optimal beamforming and power allocation design was studied for
 multiuser narrow-band systems with multiple transmit antennas in \cite{JR:Rui_zhang_power_splitting}, while the resource allocation algorithm design for wide-band SWIPT systems  was studied in \cite{CN:WCNC_WIPT}\nocite{JR:WIPT_fullpaper}--\cite{JR:Kai_bin}. In  \cite{CN:WCNC_WIPT} and \cite{JR:WIPT_fullpaper},  the energy efficiency of  multi-carrier modulation  with  SWIPT was investigated for  single user and multiuser systems, respectively. In particular, it was shown in   \cite{JR:WIPT_fullpaper} that the  energy efficiency of a communication system can be improved by integrating an energy harvester into a conventional information receiver which further motivates the deployment of SWIPT in practice.
Besides,  a power allocation algorithm was designed for the maximization of spectral efficiency of SWIPT systems employing power splitting receivers in \cite{JR:Kai_bin}. The results in \cite{Pa:Nikola}--\cite{JR:Kai_bin} reveal that  the amount of harvested energy at the receivers can be increased by increasing the transmit power of the information signals.  However, a high signal power may also lead to substantial information
leakage  due to the broadcast nature of the wireless communication channel and facilitate eavesdropping.

The notion  of  physical (PHY) layer security  in SWIPT systems has recently been pursued in \cite{JR:MOOP}--\nocite{JR:rui_zhang}\cite{JR:Kwan_secure_imperfect}.
 By exploiting multiple antennas, transmit beamforming and
artificial noise generation can be utilized  for providing communication security while guaranteeing QoS in wireless energy transfer to energy harvesting receivers.
However, the resource allocation algorithms in \cite{JR:MOOP}--\cite{JR:Kwan_secure_imperfect} were designed for a single information receiver and single-antenna eavesdroppers. The  results  in \cite{JR:MOOP}--\cite{JR:Kwan_secure_imperfect} may not be applicable to the case of multiple-antenna eavesdroppers. Besides, the works in \cite{CN:Shannon_meets_tesla}--\nocite{JR:MIMO_WIPT,JR:WIP_receiver,
JR:Rui_zhang_power_splitting,
CN:WCNC_WIPT,JR:WIPT_fullpaper,JR:Kai_bin,JR:MOOP}\cite{JR:Kwan_secure_imperfect}    did not take into account  fairness issues in transferring energy to energy harvesting receivers. Nevertheless, fairness is an essential QoS figure of merit for wireless energy transfer.

In this paper, we focus on the resource allocation algorithm design for providing fairness to energy receivers in the wireless energy transfer process  while guaranteing communication secrecy  to information receivers.
The resource allocation algorithm design is formulated as a non-convex optimization problem. In particular, we promote the dual use of artificial noise for secrecy communication and efficient wireless energy transfer provisioning.   The considered non-convex problem is solved optimally by  semidefinite programming (SDP) relaxation and our simulation results  unveil  the potential performance gain achieved by the proposed optimization framework.


\section{System Model}
\label{sect:system model}
\subsection{Notation}
We use boldface capital and lower case letters to denote matrices and vectors, respectively. $\mathbf{A}^H$, $\Tr(\mathbf{A})$, $\Rank(\mathbf{A})$, and $\det(\mathbf{A})$ represent the Hermitian transpose, trace, rank, and determinant of  matrix $\mathbf{A}$, respectively; $\lambda_{\max}(\mathbf{A})$ denotes the maximum eigenvalue of matrix $\mathbf{A}$; $\mathbf{A}\succ \zero$ and $\mathbf{A}\succeq \zero$ indicate that $\mathbf{A}$ is a positive definite and a  positive semidefinite matrix, respectively; $\mathbf{I}_N$ is the $N\times N$ identity matrix; $\mathbb{C}^{N\times M}$ denotes the set of all $N\times M$ matrices with complex entries; $\mathbb{H}^N$ denotes the set of all $N\times N$ Hermitian matrices.  The circularly symmetric complex Gaussian (CSCG) distribution is denoted by ${\cal CN}(\mathbf{m},\mathbf{\Sigma})$ with mean vector $\mathbf{m}$ and covariance matrix $\mathbf{\Sigma}$; $\sim$ indicates ``distributed as"; ${\cal E}\{\cdot\}$ denotes  statistical expectation; $\abs{\cdot}$ represents the absolute value of a complex scalar. $[x]^+$ stands for $\max\{0,x\}$.

 \begin{figure}
 \centering
\includegraphics[width=3.5in]{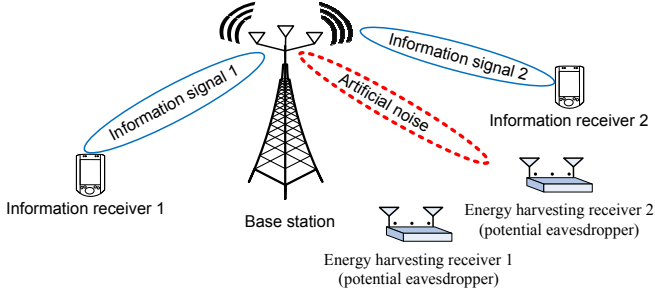}
 \caption{Downlink SWIPT communication system model with $K=2$ single-antenna information receivers and $J=2$ multiple-antenna energy harvesting receivers.}
 \label{fig:system_model}
\end{figure}
\subsection{Channel Model}
We consider a multiuser downlink communication system with SWIPT.  In particular, a base station (/transmitter) equipped with   $N_\mathrm{T}>1$ antennas serving  $K$  information receivers and $J$ energy harvesting receivers is considered. The $K$ information receivers are low  complexity single-antenna devices while the $J$ energy harvesting receivers are equipped with $N_{\mathrm{R}}\ge 1$ antennas. We assume that $N_\mathrm{T}>N_\mathrm{R}$. A possible application of the considered system model is cognitive radio. In particular, the energy harvesting receivers may be the primary users which are temporally connecting to a secondary transmitter with the intend to harvest energy from the received signals   for extending the lifetime of the primary network. In turn, the primary network may grant the secondary network spectrum usage rights for the energy supply services.  In each  time slot, $K$ independent signal streams are transmitted simultaneously to $K$  information receivers. To prevent the $J$ energy harvesting receivers (potential eavesdroppers) from eavesdropping the  information for the $ K$ intended information receivers,  artificial noise is transmitted concurrently with the information signals  for interfering the reception of the energy harvesting receivers. As a result, the transmitted signal vector, $\mathbf{x}\in\mathbb{C}^{N_{\mathrm{T}}\times 1}$, consists of the $K$ desired information signals and artificial noise, and  can be expressed as
\begin{eqnarray}
\mathbf{x}&=&\underbrace{\sum_{k=1}^K\mathbf{w}_k s_k}_{\mbox{desired signals}}+\underbrace{\mathbf{v}}_{\mbox{artifical noise}},
\end{eqnarray}
where $s_k\in\mathbb{C}$ and $\mathbf{w}_k\in\mathbb{C}^{N_{\mathrm{T}}\times1}$  are the symbol  and  the beamforming vector intended for information receiver $k$, respectively. Variable  $\mathbf{v}\in\mathbb{C}^{N_{\mathrm{T}}\times1}$ is the artificial noise vector  generated by the transmitter and  is modelled  as $\mathbf{v}\sim\cal{CN}(\zero,\mathbf{V})$ with zero mean and covariance matrix $\mathbf{V}$, $\mathbf{V}\in\mathbb{H}^{N_{\mathrm{T}}}, \mathbf{V}\succeq \zero$. Without loss of generality we assume that ${\cal E}\{\abs{s_k}^2\}=1,\forall k\in\{1,\ldots,K\}$.

In this paper, we focus on a frequency flat slow fading channel. The received signals at the information receivers and the energy harvesting receivers are given by
\begin{eqnarray}
y_{k}&=&\mathbf{h}_k^H\mathbf{x}+n_k,\,\,  \forall k\in\{1,\dots,K\},\,\mbox{and}\\
\mathbf{y}_{\mathrm{ER}_j}&=&\mathbf{G}_j^H\mathbf{x}+\mathbf{n}_{\mathrm{ER}_j},\,\,  \forall j\in\{1,\dots,J\},
\end{eqnarray}
respectively. The channel vector between the transmitter and  desired receiver $k$ is denoted by $\mathbf{h}_k\in\mathbb{C}^{N_{\mathrm{T}}\times1}$. The channel  matrix between the transmitter and  energy harvesting receiver $j$ is denoted by $\mathbf{G}_j\in\mathbb{C}^{N_{\mathrm{T}}\times N_{\mathrm{R}}}$. The channel vectors and matrices capture the joint effects of multipath fading and path loss. $n_k\sim{\cal CN}(0,\sigma_{\mathrm{s}}^2)$ and $\mathbf{n}_{\mathrm{ER}_j}\sim{\cal CN}(\zero,\sigma_{\mathrm{s}}^2\mathbf{I}_{N_{\mathrm{R}}})$ are additive white Gaussian noises (AWGN) caused by the thermal noises in the antennas of the information receivers and the energy harvesting receivers, respectively. $\sigma_{\mathrm{s}}^2$ denotes the noise power at the receiver.

\section{Resource Allocation Algorithm Design}
\label{sect:forumlation}

\subsection{Energy Harvesting}
\label{subsect:harvested energy}
In the considered system, the energy harvesting receivers scavenge power\footnote{ In this paper, we study the algorithm design for a normalized unit energy, i.e., Joule-per-second. Thus, the terms ``energy" and ``power" are used interchangeably.} from the RF. The total amount of power harvested by energy harvesting receiver $j$ is given by
\begin{eqnarray}
E_{\mathrm{ER}_j}&=&\eta_j\Tr\Big(\mathbf{G}_j^H\big(\sum_{k=1}^K\mathbf{w}_k\mathbf{w}_k^H+\mathbf{V}\big)\mathbf{G}_j\Big),
\end{eqnarray}
where $0\le\eta_j\le 1$ is a constant which denotes the energy conversion efficiency of energy harvesting receiver $j$.

\subsection{Channel Capacity and Secrecy Capacity}
\label{subsect:Instaneous_Mutual_information}
 Given perfect channel state information (CSI) at the
receiver, the  channel capacity (bit/s/Hz) between the transmitter and information receiver $k$ is given by
\begin{eqnarray}
C_{k}\hspace*{-1mm}&=&\hspace*{-1mm}\log_2(1+\Gamma_{k}),\quad \mbox{where}\\
\Gamma_{k}\hspace*{-1mm}&=&\hspace*{-1mm}\frac{\abs{\mathbf{h}_k^H\mathbf{w}_k}^2}{\sum\limits_
{\substack{m\neq k}}^K\abs{\mathbf{h}_k^H\mathbf{w}_m}^2+\Tr(\mathbf{H}_k\mathbf{V})+\sigma_{\mathrm{s}}^2}
\end{eqnarray}
is the receive signal-to-interference-plus-noise ratio (SINR) at information receiver $k$ and $\mathbf{H}_k=\mathbf{h}_k\mathbf{h}_k^H$.

On the other hand,  we focus on  an unfavourable scenario for the
decoding capability of the energy harvesting receivers for providing communication security to the information receivers. Specifically,  we assume that energy harvesting receiver $j$ performs interference cancellation  to remove  all multiuser interference and eavesdrops the message intended for information receiver $k$. Therefore, the channel capacity between the transmitter and energy harvesting receiver $j$ for decoding the signal of information receiver $k$ can be represented as
\begin{eqnarray}\label{eqn:Capacity_eve}
\hspace*{-6mm}C_{\mathrm{ER}_{j}}^k\hspace*{-2mm}&=&\hspace*{-2mm}\log_2\det(\mathbf{I}_{N_{\mathrm{R}}}\hspace*{-0.5mm}+\hspace*{-0.5mm}
\mathbf{Q}_{j}^{-1}
\mathbf{G}_j^H\mathbf{w}_k\mathbf{w}_k^H\mathbf{G}_j),\\
\notag\mathbf{Q}_{j}\hspace*{-2mm}&=&\hspace*{-2mm}\mathbf{G}_j^H\mathbf{V}\mathbf{G}_j+
\sigma_{\mathrm{s}}^2\mathbf{I}_{N_{\mathrm{R}}}\hspace*{-0.5mm}\succ\hspace*{-0.5mm} \zero,
\end{eqnarray}
where $\mathbf{Q}_{j}$ is the  interference-plus-noise covariance matrix for energy harvesting receiver $j$ assuming the worst case for  communication secrecy. Thus, the  achievable secrecy capacity of information receiver $k$ is given by
\begin{eqnarray}\label{eqn:secure_cap}
C_{\mathrm{sec}_k}&=&\Big[C_{k}-\underset{\forall j}{\max}\,\{ C_{\mathrm{ER}_{j}}^k\}\Big]^+.
\end{eqnarray}

\subsection{Optimization Problem Formulation}
\label{sect:cross-Layer_formulation}
The system objective is to maximize the minimum  harvested power among all  energy harvesting receivers while providing QoS for communication security. The resource allocation algorithm design  is formulated as an optimization problem which is given by
\begin{eqnarray}\label{eqn:cross-layer}\notag
&&\hspace*{-10mm} \underset{ \mathbf{V}\in \mathbb{H}^{N_{\mathrm{T}}},\mathbf{w}_k}{\maxo}\,\, \min_{j\in\{1,\ldots,J\}}\Big\{\eta_j\Tr\Big(\mathbf{G}_j^H\big(\sum_{k=1}^K\mathbf{w}_k\mathbf{w}_k^H+\mathbf{V}\big)\mathbf{G}_j\Big)\Big\}\\
\notag \mbox{s.t.} &&\hspace*{-1mm}\mbox{C1: }\Gamma_{k}\ge \Gamma_{\mathrm{req}_k},\,\, \forall k, \notag\\
&&\hspace*{-1mm}\mbox{C2: }C_{\mathrm{ER}_j}^{k}\le R_{\mathrm{ER}_{j,k}},\,\, \forall k,\forall j,\notag\\
&&\hspace*{-1mm}\mbox{C3: }\Tr(\mathbf{V})+\sum_{k=1}^K\norm{\mathbf{w}_k}^2\le P_{\max},\,\,\mbox{C4:}\,\, \mathbf{V}\succeq \zero.
\end{eqnarray}
Constraint C1 indicates that the receive SINR at information receiver $k$ is required to be larger than a given threshold, $\Gamma_{\mathrm{req}_k}>0$, for guaranteeing reliable communication. The upper limit $R_{\mathrm{ER}_{j,k}}>0$ in C2 is imposed  to restrict the channel capacity of energy harvesting receiver $j$ if it  attempts to decode the message of information receiver $k,\forall k$.   Constant $P_{\max}$  in constraint C3 limits the radiated power from the transmitter accounting for the power budget of the transmitter. Constraint C4 and $\mathbf{V}\in \mathbb{H}^{N_{\mathrm{T}}}$ ensure that the covariance matrix $\mathbf{V}$ is  a positive semidefinite Hermitian matrix.

\subsection{Optimization Solution}\label{sect:solution}
Problem (\ref{eqn:cross-layer}) is a non-convex optimization problem. In particular, the non-convexity arises from    constraints C1 and C2. To overcome the non-convexity,  we first propose the following proposition and then  recast the considered problem into a convex optimization problem using SDP relaxation.

\begin{proposition}\label{prop:relaxed_c2} For $R_{\mathrm{ER}_{j,k}}> 0,\forall j,k$, the following implication on constraint C2 holds:
\begin{eqnarray}\label{eqn:det_to_matrix}
\mbox{C2}\Rightarrow\overline{\mbox{C2}}\mbox{: } \mathbf{G}_j^H\mathbf{W}_k\mathbf{G}_j\preceq \alpha_{\mathrm{ER}_{j,k}} \mathbf{Q}_{j},\,\, \forall j,k,
\end{eqnarray}
\end{proposition}
where $\mathbf{W}_k=\mathbf{w}_k\mathbf{w}_k^H$ and $\alpha_{\mathrm{ER}_{j,k}}=2^{R_{\mathrm{ER}_{j,k}}}-1$ is an auxiliary constant.   $\overline{\mbox{C2}}$ is a linear matrix inequality (LMI) constraint. We note that constraints $\overline{\mbox{C2}}$ and ${\mbox{C2}}$ are equivalent if $\Rank(\mathbf{W}_k)\le 1,\forall k$.

\,\,\emph{Proof:} Please refer to the Appendix for the proof.\qed

Now, we apply Proposition \ref{prop:relaxed_c2} to (\ref{eqn:cross-layer}) and replace constraint $\mbox{C2}$ with constraint $\overline{\mbox{C2}}$.  By  setting $\mathbf{W}_k\in\mathbb{H}^{N_{\mathrm{T}}},\forall k$, $\mathbf{W}_k=\mathbf{w}_k\mathbf{w}_k^H$, and $\Rank(\mathbf{W}_k)\le 1,\,\, \forall k$, we can rewrite the
optimization problem  in its hypograph form:
\begin{eqnarray}\label{eqn:rank_one}
&&\hspace*{15mm} \underset{\mathbf{W}_k, \mathbf{V}\in \mathbb{H}^{N_{\mathrm{T}}}, \tau}{\maxo}\,\,\,\, \tau\\
\mbox{s.t.} &&\hspace*{-5mm}\mbox{C1: }\frac{\Tr(\mathbf{H}_k\mathbf{W}_k)}{\Gamma_{\mathrm{req}_k}}-\Tr\Big(\mathbf{H}_k\big(\sum\limits_
{\substack{m\neq k}}^K\mathbf{W}_m+\mathbf{V}\big)\Big)\ge\sigma_{\mathrm{s}}^2,\,\, \forall k, \notag\\
&&\hspace*{-5mm}\overline{\mbox{C2}}\mbox{: } \notag\mathbf{G}_j^H\mathbf{W}_k\mathbf{G}_j\preceq \alpha_{\mathrm{ER}_{j,k}}\mathbf{Q}_{j},\,\, \forall j,k,\notag\\
&&\hspace*{-5mm}\mbox{C3: }\notag\Tr\Big(\mathbf{V}+\sum_{k=1}^K\mathbf{W}_k\Big)\le P_{\max},\notag\\
&&\hspace*{-8mm}\mbox{C4:}\,\,
\notag\mathbf{V}\succeq \zero,\,\,\mbox{C5:}\,\,  \mathbf{W}_k\succeq \zero, \forall k,\,\, \mbox{C6:}\,\, \Rank(\mathbf{W}_k)\le 1,\,\, \forall k,\\
&&\hspace*{-9mm}\mbox{C7:}\,\,  \eta_j\Tr\Big(\mathbf{G}_j^H\big(\sum_{k=1}^K\mathbf{W}_k+\mathbf{V}\big)\mathbf{G}_j\Big)\ge \tau,\forall j\in\{1,\ldots,J\},\notag
\end{eqnarray}
where $\tau$ is an auxiliary optimization variable.  Then, we adopt  SDP relaxation by removing constraint $\mbox{C6: }\Rank(\mathbf{W}_k)\le 1$  from the problem formulation which results in a convex SDP problem.  The relaxed SDP problem formulation of (\ref{eqn:rank_one}) is given by
\begin{eqnarray}
\label{eqn:sdp_relaxation}&&\hspace*{-5mm}\underset{\mathbf{W}_k, \mathbf{V}\in \mathbb{H}^{N_{\mathrm{T}}}, \tau}{\maxo}\,\,\,\, \tau\notag\\
&&\hspace*{-15mm}\mbox{s.t. } \mbox{C1},\,\overline{\mbox{C2}},\,\mbox{C3},\,\mbox{C4},\,\mbox{C5},\,\mbox{C7}.
\end{eqnarray}
We note that the relaxed problem in (\ref{eqn:sdp_relaxation}) can be solved efficiently by numerical solvers such as CVX \cite{website:CVX}. If the obtained solution for (\ref{eqn:sdp_relaxation}), $\mathbf{W}_k$,  admits a rank-one matrix, then the problems in (\ref{eqn:cross-layer}), (\ref{eqn:rank_one}), and (\ref{eqn:sdp_relaxation}) share the same optimal solution and the same optimal objective value.  In the following, we investigate if $\Rank(\mathbf{W}_k)=1$ holds for the solution of (\ref{eqn:sdp_relaxation}).

It can be shown that the problem in  (\ref{eqn:rank_one}) satisfies Slater's constraint qualification. Thus, strong duality holds and solving the dual problem is equivalent to solving the primal problem. The Lagrangian of (\ref{eqn:sdp_relaxation}) is expressed as:
\begin{eqnarray}
&&\hspace*{-5mm}{\cal L}\Big(\mathbf{W}_k,\mathbf{V},\tau,\mathbf{Z}_k,\mathbf{Y},\mathbf{X}_{j,k},\delta_k,\lambda,\beta_j\Big)\\
=&&\hspace*{-5mm}\tau-\sum_{j=1}^J\beta_j\Big[ \tau- \eta_j\Tr\Big(\mathbf{G}_j^H\big(\sum_{k=1}^K\mathbf{W}_k\hspace*{-0.5mm}+\hspace*{-0.5mm}\mathbf{V}\big)\mathbf{G}_j\Big) \Big]\notag\\
-&&\hspace*{-5mm}\sum_{k=1}^K\delta_k\Big[-\frac{\Tr(\mathbf{H}_k\mathbf{W}_k)}{\Gamma_{\mathrm{req}_k}}
+\hspace*{-0.5mm}
\Tr\Big(\mathbf{H}_k\big(\sum\limits_
{\substack{m\neq k}}^K\mathbf{W}_m\hspace*{-0.5mm}+\hspace*{-0.5mm}\mathbf{V}\big)\Big)\hspace*{-0.5mm}+\hspace*{-0.5mm}\sigma_{\mathrm{s}}^2\Big]\notag\\
+&&\hspace*{-5mm}\Tr(\mathbf{Y}\mathbf{V})\hspace*{-0.5mm}+\hspace*{-0.5mm}
\sum_{k=1}^K\Tr(\mathbf{Z}_k\mathbf{W}_k)
\hspace*{-0.5mm}-\hspace*{-0.5mm}\lambda\Big[\hspace*{-0.5mm}\Tr\Big(\mathbf{V}\hspace*{-0.5mm}+\hspace*{-0.5mm}\sum_{k=1}^K\mathbf{W}_k\Big)- P_{\max}\hspace*{-0.5mm}\Big]\notag\\
-&&\hspace*{-5mm}\sum_{j=1}^J\sum_{k=1}^K\Tr\Big\{\mathbf{X}_{j,k}\Big[\mathbf{G}_j^H\mathbf{W}_k\mathbf{G}_j-
\alpha_{\mathrm{ER}_{j,k}}\mathbf{Q}_{j}\Big] \Big\}\notag,
\end{eqnarray}
where $\mathbf{X}_{j,k}$, $\mathbf{Y}$, and $\mathbf{Z}_k$ are the dual variable matrices for constraints $\overline{\mbox{C2}}$, C4, and C5, respectively.  $\delta_k$,  $\lambda$, and $\beta_j$ are the scalar dual variables for constraints  C1, C3, and C7 respectively. Let $\mathbf{B}_k=\lambda\mathbf{I}_{N_\mathrm{T}}+\sum_{j=1}^J\mathbf{G}_j(\beta_j\mathbf{I}_{N_\mathrm{R}}-\mathbf{X}_{j,k})\mathbf{G}_j^H+\sum_{m\ne k}\mathbf{H}_m\delta_m$, and $\Rank(\mathbf{B}_k)=r_k$. Besides, we denote the orthonormal basis of the null space of $\mathbf{B}_k$ as $\mathbf{\mathbf{\Upsilon}}_k\in\mathbb{C}^{N_{\mathrm{T}}\times (N_{\mathrm{T}}-r_{k})}$, and ${\boldsymbol \phi}_{\nu_{k}}\in \mathbb{C}^{N_{\mathrm{T}}\times 1}$, where $1\le \nu_{k}\le N_{\mathrm{T}}-r_{k}$, denotes the $\nu_{k}$-th column  of $\mathbf{\Upsilon}_{k}$. Hence, $\mathbf{B}_{k}\mathbf{\Upsilon}_{k}=\zero$ and $\Rank(\mathbf{\Upsilon}_{k})=N_{\mathrm{T}}-r_{k}$. Also, it can be shown that $\mathbf{H}_k\mathbf{\Upsilon}_{k}=\zero$ holds for the optimal solution.

Now, we introduce the following  theorem for revealing the tightness of the SDP relaxation adopted in  (\ref{eqn:sdp_relaxation}).
\begin{Thm}\label{thm:rankone_condition} Suppose the optimal solution of (\ref{eqn:sdp_relaxation}) is denoted by $\{\mathbf{W}_k^*,\mathbf{V}^*,\tau^*\}$, ${\Gamma}_{\mathrm{req}_k}>0$, and $R_{\mathrm{ER}_{j,k}}>0$. The optimal solution, $\mathbf{W}_k^*$, can be expressed as
\begin{eqnarray}\label{eqn:general_structure}
\mathbf{W}^*_{k}=\sum_{\nu_{k}=1}^{N_{\mathrm{T}}-r_{k}}\psi_{\nu_{k}}{\boldsymbol \phi}_{\nu_{k}}  {\boldsymbol \phi}_{\nu_{k}} ^H  + \underbrace{f_{k}\mathbf{u}_{k}\mathbf{u}^H_{k}}_{\mbox{rank-one}},
\end{eqnarray}
where $\psi_{\nu_{k}}\ge0, \forall \nu_{k}\in\{1,\ldots,N_{\mathrm{T}}-r_{k}\},$ and $f_{k}>0$ are positive scalars and $\mathbf{u}_{k}\in \mathbb{C}^{N_{\mathrm{T}}\times 1}$, $\norm{\mathbf{u}_{k}}=1$, satisfies $\mathbf{u}^H_{k}\mathbf{\Upsilon}_{k}=\zero$. If  $\exists k:\Rank(\mathbf{W}^*_k)>1$, i.e., $\psi_{\nu_{k}}>0$, then we can construct another solution of (\ref{eqn:sdp_relaxation}), denoted by  $\{\mathbf{\widetilde  W}_k ,\mathbf{\widetilde  V},\widetilde \tau\}$, which not only achieves the same objective value as $\{\mathbf{W}_k^*,\mathbf{V}^*,\tau^*\}$, but also admits a rank-one matrix, i.e.,  $\Rank(\mathbf{\widetilde W}_k)=1,\forall k$. The new optimal solution is given as
\begin{eqnarray}\label{eqn:rank-one-structure}\mathbf{\widetilde W}_{k}\hspace*{-2mm}&=&\hspace*{-2mm}f_{k}\mathbf{u}_{k}\mathbf{u}^H_{k}=\mathbf{W}^*_{k}-\sum_{\nu_{k}=1}^{N_{\mathrm{T}}-r_{k}} \psi_{\nu_{k}} {\boldsymbol \phi}_{\nu_{k}} {\boldsymbol \phi}_{\nu_{k}}^H,\\
 \mathbf{\widetilde V}\hspace*{-2mm}&=&\hspace*{-2mm}\mathbf{ V}^*+\sum_{\nu_{k}=1}^{N_{\mathrm{T}}-r_{k}} \psi_{\nu_{k}} {\boldsymbol \phi}_{\nu_{k}} {\boldsymbol \phi}_{\nu_{k}}^H,\quad
\widetilde\tau=\tau^*,\notag
\end{eqnarray}
with $\Rank(\mathbf{\widetilde W}_{k})=1,\forall k\in\{1,\ldots,K\}$, where $f_{k}$ and $ \psi_{\nu_{k}}$ can be easily
found  by substituting the variables in (\ref{eqn:rank-one-structure}) into
the relaxed version of (\ref{eqn:sdp_relaxation}) and solving the resulting convex
optimization problem for $f_{k}$ and $ \psi_{\nu_{k}}$.
\end{Thm}
\emph{\quad Proof: } The  proof of Theorem 1 is similar to the proof of  \cite[Proposition 4.1]{JR:rui_zhang} and omitted here for brevity.\qed

By combining the results of Proposition 1 and Theorem 1, the global optimal solution of (\ref{eqn:cross-layer}) can be obtained by solving (\ref{eqn:sdp_relaxation}) even though SDP relaxation is applied.

\begin{table}[t]\caption{System parameters}\label{tab:feedback} \centering
\begin{tabular}{|L|M|}\hline

Carrier center frequency & 915 MHz  \\
  \hline
Small-scale fading distribution & Rician fading with Rician \hspace*{-2.5mm} factor $3$ dB \\

    \hline
    Total noise variance, $\sigma_{\mathrm{s}}^2$ &  \mbox{$-23$ dBm}   \\
    \hline
    Transmit power budget, $P_{\max}$ &  $46$ \mbox{dBm}   \\
        \hline
    Number of receive antennas at each ER, $N_{\mathrm{R}}$ &  $2$    \\
        \hline
    Receive antenna gain &  $6$ dB  \\
    \hline
         Max. tolerable channel capacity at ERs,  \hspace*{-1.5mm}$R_{\mathrm{ER}_{j,k}}$ & $1$ bit/s/Hz \\
                \hline
                                         RF energy to electrical energy conversion \hspace*{-1.5mm}efficiency for ER $j$, $\eta_j$ & $0.5$  \\
 \hline

\end{tabular}\vspace*{-5mm}
\end{table}
\section{Results}
In this section, we study the system performance of the proposed resource allocation scheme via simulation.  There are $K=3$ information receivers and $J=2$ energy harvesting receivers,  which are uniformly distributed in the range between a reference distance of $2$ meters and a maximum distance of $50$ meters. The detailed simulation parameters can be found in Table \ref{tab:feedback}. We assume that  all information receivers require the same minimum SINR, i.e., $\Gamma_{\mathrm{req}_k}=\Gamma_{\mathrm{req}},\forall k\in\{1,\ldots,K\}$, for illustration. Besides, we solve the optimization problem in (\ref{eqn:cross-layer}) and obtain the  average system performance by averaging over  different channel realizations.

\subsection{Average Total Harvested Power}
\begin{figure}[t]
 \centering
\includegraphics[width=3.5in]{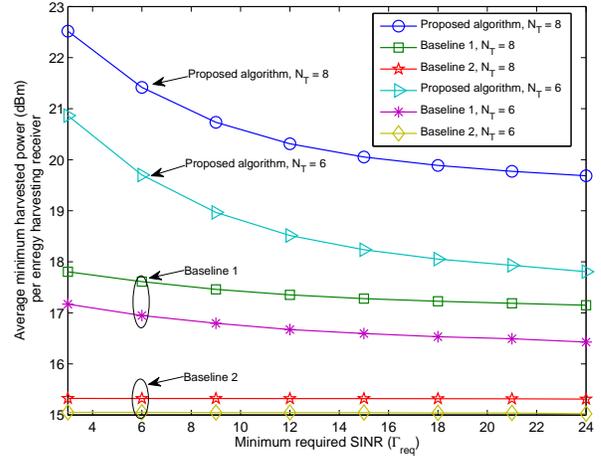}\vspace*{-1mm}
 \caption{Average minimum harvested power  (dBm)  per energy harvesting receiver versus the minimum required SINR  of the information receivers, $\Gamma_{\mathrm{req}}$.}
 \label{fig:hp_SINR}\vspace*{-5mm}
\end{figure}

In Figure \ref{fig:hp_SINR}, we study the average minimum harvested power per energy harvesting receiver of the optimal scheme versus  the minimum required SINR,  $\Gamma_{\mathrm{req}}$, for different numbers of transmit antennas  and different resource allocation schemes. It can be observed that the minimum harvested power decreases with $\Gamma_{\mathrm{req}}$. Indeed, for satisfying a more stringent SINR requirement, the transmitter is forced to  allocate a higher power to the information signal and to steer the  direction of transmission towards the information receivers. Besides, some degrees of freedom have to be sacrificed  for  artificial noise transmission for reducing the interference to the information receivers.    This leads to a smaller amount of RF energy for energy harvesting.  On the other hand,   when the number of antennas increase from $N_{\mathrm{T}}=6$ to $N_{\mathrm{T}}=8$,  a significant energy harvesting gain can be achieved by the proposed optimal scheme. In fact, the degrees of freedom for resource allocation increase with the number of transmit antennas, which enables a more power efficient energy transfer to the energy harvesting receivers.

For comparison, we also show  the performance of two simple suboptimal baseline schemes for $N_{\mathrm{T}}>K$. For baseline scheme 1, the artificial noise is transmitted into the null space spanned by the channel of the $K$ information receivers. In other words, the artificial noise does not interfere with the desired information receivers.     Then, we maximize the minimum harvested power at the energy receivers by optimizing  $\mathbf{W}_k$ and the power of artificial noise subject to the same constraints as in (\ref{eqn:sdp_relaxation}). We note that the obtained  $\mathbf{W}_k$ for baseline scheme 1 may not be a rank-one matrix. As for baseline scheme 2, it shares the same resource allocation policy  as baseline scheme 1 except that the direction of beamforming matrix, $\mathbf{W}_k$, is fixed. In particular,  we calculate the null space of $\mathbf{\widetilde h}_{-k}\mathbf{\widetilde h}_{-k}^H$ for desired information receiver $k$ where $\mathbf{\widetilde h}_{-k}=[\mathbf{h}_1\,\ldots\,\mathbf{h}_{k-1}\,\mathbf{h}_{k+1}\,\ldots\,\mathbf{h}_K]$. Then, we project the vector $\mathbf{h}_{k}$ onto the null space of $\mathbf{\widetilde h}_{-k}\mathbf{\widetilde h}_{-k}^H$ and use the resulting vector as the direction of beamforming vector $\mathbf{w}_k$. We note that $\mathbf{W}_k$ in baseline scheme 2 is a rank-one matrix by construction. It can be observed in Figure \ref{fig:hp_SINR} that for the proposed optimal scheme, the energy harvesting receivers are able to harvest more energy compared to the two baseline schemes, due to the joint optimization of $\mathbf{W}_k$ and $\mathbf{V}$. Besides, the performance gain of  the optimal scheme over the two baseline schemes  is further enlarged for an increasing number of transmit antennas  $N_{\mathrm{T}}$. This can be explained by the fact that the optimal scheme can fully utilize the degrees of freedom offered by the system for resource allocation. In contrast,  although multiuser and artificial noise interference are eliminated at the information receivers  for  baseline scheme 1 and baseline scheme 2, respectively, the degrees of freedom for resource allocation in the baseline schemes are reduced which results in a lower harvested energy at the energy harvesting receivers.

\subsection{Secrecy Capacity}
\begin{figure}[t]
 \centering
\includegraphics[width=3.5in]{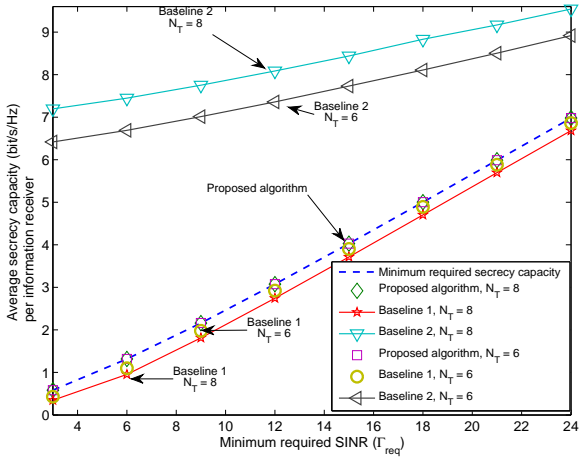}\vspace*{-1mm}
 \caption{Average secrecy capacity  (bit/s/Hz) per information receiver versus the minimum required SINR  of the information receivers, $\Gamma_{\mathrm{req}}$.}
 \label{fig:cap_SINR}\vspace*{-5mm}
\end{figure}

Figure \ref{fig:cap_SINR} illustrates the average secrecy capacity per information receiver versus the minimum required SINR $\Gamma_{\mathrm{req}}$ of the information receivers for different numbers of transmit antennas and different resource allocation schemes.  It can be observed that the average system secrecy capacity, i.e., $C_{\mathrm{sec}_k}$, is a non-decreasing function with  respect to $\Gamma_{\mathrm{req}}$. In fact,  the channel capacity of energy harvesting receiver $j$ for decoding information receiver $k$ is constrained to be less than $R_{\mathrm{ER}_{j,k}}= 1$ bit/s/Hz, cf. Table I. Besides, we note that baseline scheme 1 is unable to meet the minimum required of secrecy capacity as specified by constraints C1 and C2. In other words,
there are time instants for baseline 1 such that $\Rank(\mathbf{W}_k)>1$. Thereby, although baseline scheme 1 satisfies constraint $\overline{\mbox{C2}}$, unlike the optimal scheme, it does not necessarily  satisfy constraint C2. On the other hand, both the proposed algorithm and baseline  scheme 2 are able to meet the minimum required secrecy capacity due to the rank-one solution for beamforming  matrix $\mathbf{W}_k$.  However, the exceedingly large secrecy capacity achieved by baseline scheme 2 comes at the expense of a smaller harvested power compared to the proposed optimal scheme, cf. Figure \ref{fig:hp_SINR}.

\section{Conclusions}\label{sect:conclusion}
In this paper, we  studied the resource allocation algorithm design for SWIPT.
 The algorithm design was formulated as a non-convex  optimization problem to ensure the max-min fairness in energy transfer to the energy harvesting receivers.  The proposed problem formulation enabled the dual use of artificial noise for   efficient energy transfer and  secure communication.  SDP relaxation was adopted to obtain the optimal solution of the considered non-convex optimization problem.  Simulation results unveiled the potential gain in  harvested energy of our proposed optimal resource allocation scheme compared to baseline schemes.

\section*{Appendix-Proof of Proposition \ref{prop:relaxed_c2}}
We start the proof by rewriting constraint C2 as
\begin{eqnarray}\label{eqn:det_ineq}\hspace*{-8mm}
\mbox{C2: }\log_2\det(\mathbf{I}_{N_{\mathrm{R}}}+\mathbf{Q}_{j}^{-1}\mathbf{G}_j^H\mathbf{W}_k\mathbf{G}_j)\hspace*{-3mm}&\le&\hspace*{-3mm}R_{\mathrm{ER}_{j,k}}\\
\hspace*{-6mm}\Longleftrightarrow \det(\mathbf{I}_{N_{\mathrm{R}}}+\mathbf{Q}_{j}^{-1/2}\mathbf{G}_j^H\mathbf{W}_k\mathbf{G}_j\mathbf{Q}_{j}^{-1/2})
\hspace*{-3mm}&\le&\hspace*{-3mm}1+\alpha_{\mathrm{ER}_{j,k}}\label{eqn:det_ineq2}.
\end{eqnarray}
 Then, we propose a lower bound on the left hand side of (\ref{eqn:det_ineq2}) by introducing the following lemma.
\begin{Lem}\label{lemma:det_trace} For any square matrix $\mathbf{A}\succeq \zero$, we have the following inequality \cite{JR:AN_MISO_secrecy}:
\begin{eqnarray}
\det(\mathbf{I}+\mathbf{A})\ge 1+\Tr(\mathbf{A}),
\end{eqnarray}
\end{Lem}
where the equality holds if and only if $\Rank(\mathbf{A})\le 1$.

Exploiting Lemma \ref{lemma:det_trace}, the left hand side of (\ref{eqn:det_ineq2}) is bounded below by
\begin{eqnarray}\hspace*{-10mm}&&\det(\mathbf{I}_{N_{\mathrm{R}}}+\mathbf{Q}_{j}^{-1/2}\mathbf{G}_j^H\mathbf{W}_k\mathbf{G}_j\mathbf{Q}_{j}^{-1/2}) \notag\\
\hspace*{-10mm}&{\ge}& 1+\Tr(\mathbf{Q}_{j}^{-1/2}\mathbf{G}_j^H\mathbf{W}_k\mathbf{G}_j\mathbf{Q}_{j}^{-1/2}).\label{eqn:trace_ineq3}
\end{eqnarray}

Subsequently, by combining equations (\ref{eqn:det_ineq}), (\ref{eqn:det_ineq2}),  and (\ref{eqn:trace_ineq3}), we have the following implications:
\begin{subequations}
\begin{eqnarray}\notag
\hspace*{-5mm}&&\mbox{(\ref{eqn:det_ineq})} \Longleftrightarrow \mbox{(\ref{eqn:det_ineq2})}\notag\\
\hspace*{-5mm}&\Longrightarrow& \hspace*{-2mm}\Tr(\mathbf{Q}_{j}^{-1/2}\mathbf{G}_j^H\mathbf{W}_k\mathbf{G}_j\mathbf{Q}_{j}^{-1/2})\le \alpha_{\mathrm{ER}_{j,k}}\\
\hspace*{-5mm}&\Longrightarrow &\hspace*{-2mm} \lambda_{\max}(\mathbf{Q}_{j}^{-1/2}\mathbf{G}_j^H\mathbf{W}_k\mathbf{G}_j\mathbf{Q}_{j}^{-1/2})\le \alpha_{\mathrm{ER}_{j,k}} \\
\hspace*{-5mm}&\Longleftrightarrow&\hspace*{-2mm}\mathbf{Q}_{j}^{-1/2}\mathbf{G}_j^H\mathbf{W}_k\mathbf{G}_j\mathbf{Q}_{j}^{-1/2} \preceq \alpha_{\mathrm{ER}_{j,k}}\mathbf{I}_{N_{\mathrm{R}}}\\
\hspace*{-5mm}&\Longleftrightarrow &\hspace*{-2mm} \mathbf{G}_j^H\mathbf{W}_k\mathbf{G}_j\preceq \alpha_{\mathrm{ER}_{j,k}} \mathbf{Q}_{j}. \label{eqn:trace_final}
\end{eqnarray}
\end{subequations}
 We note that  equations (\ref{eqn:det_ineq}) and (\ref{eqn:trace_final}) are equivalent, i.e., $\mbox{C2}\Leftrightarrow\overline{\mbox{C2}}$, when $\Rank(\mathbf{W}_k)= 1,\forall k$.

\end{document}